\documentclass[12pt]{iopart}
\usepackage{iopams}
\usepackage[dvips]{color}

\usepackage{graphicx}
\usepackage{multirow}
\usepackage{booktabs}

\newcommand{\ped}[1]{\ensuremath{_{\rm #1}}}
\newcommand{\apex}[1]{\ensuremath{^{\rm #1}}}
\definecolor{awesome}{rgb}{1.0, 0.13, 0.32}
\begin{document}
\title{Resistivity in Co-doped Ba-122: comparison of thin films and single crystals}
\author{G.A.~Ummarino, Sara~Galasso, Paola~Pecchio, D.~Daghero, R.S.~Gonnelli}
\ead{giovanni.ummarino@infm.polito.it}
\address{Istituto di Ingegneria e Fisica dei Materiali,
Dipartimento di Scienza Applicata e Tecnologia, Politecnico di
Torino, Corso Duca degli Abruzzi 24, 10129 Torino, Italy}
\author{F.~Kurth, K.~Iida, B.~Holzapfel}
\address{Leibniz-Institut f\"ur Festk\"orper-und Werkstoffforschung (IFW) Dresden, P.O.Box 270116, 01171 Dresden, Germany}
\begin{abstract}
The temperature dependence of the resistivity of epitaxial Ba(Fe$_{1-x}$Co$_x$)$_2$As$_2$ thin films (with nominal doping $x=0.08, 0.1$ and $0.15$) has been analyzed and compared with analogous measurements on single crystals taken from literature. The $\rho(T)$ of thin films looks different from that of single crystals, even when the cobalt content is the same. All $\rho(T)$ curves can be fitted by considering an effective two-band model (with holes and electrons bands) in which the electrons are more strongly coupled with the bosons (spin fluctuations) than holes, while the effect of impurities is mainly concentrated in the hole band. Within this model the mediating boson has the same characteristic energy in single crystals and thin films, but the shape of the transport spectral function at low energy has to be very different, leading to a ``hardening'' of the electron-boson spectral function in thin films, associated with the strain induced by the substrate.
\end{abstract}
\pacs{74.70.Xa, 74.25.F, 74.20.Mn, 74.20.-z}


%
\maketitle
\section{INTRODUCTION}
In recent years a huge effort has been devoted to the investigation of all the physical aspects of iron pnictides. These materials show very interesting properties such as the high superconducting transition temperature~\cite{ChenTc}, the proximity of the magnetic and the superconducting states and the presence of complex band structures resulting in multi-band Fermi surfaces~\cite{SinghRev}. The electronic structure in general consists of hole pockets centered at the $\Gamma$ point and electron pockets at the corner of the Brillouin zone. In particular, it has been proved that for the so-called ``122" compounds of general formula AeFe$_2$As$_2$ [Ae: alkali earth elements] the size and even the dimension of these pockets drastically change with doping~\cite{Sekiba, Brouet}.

Due to the variety and complexity of these compounds many questions about the symmetry of the order parameters, the coupling mechanism and the effects of the impurities are still open~\cite{MazinSpm,Kuroki}.

The situation becomes even more complex when a thin film is considered. In this case many differences with respect to the single crystals can appear also in the fundamental properties of the materials. Therefore a deeper investigation in this direction is indispensable~\cite{FeSe1,FeSe2}.

Nowadays very high-quality thin films are available and this allows using them for fundamental studies as well as for more technological applications~\cite{Intr1,Intr2,Intr3,Intr4,Intr5}. In particular, Co-doped BaFe$_2$As$_2$ films of high quality can be readily fabricated by pulsed laser deposition (PLD) and are thus one of the most suitable materials for these purposes. The investigation of the effects of the substrate in thin films is thus a hot topic and a particular attention has been paid to the strain-dependent critical temperature~\cite{Engelmann,KazuTcStrain}.

In this article we present the study of the resistivity measured on Ba(Fe$_{1-x}$Co$_x$)$_2$As$_2$ thin films grown on CaF$_2$ substrates and with different nominal cobalt content (namely 8\%, 10\% and 15\%) in comparison with analogous measurements on single crystals taken from literature~\cite{MisureSC}. We use a model containing two different kinds of carriers (holes and electrons) as suggested by the presence of several sheets in the Fermi surface and by the tendency of the $\rho(T)$ curves to saturate at high temperature, as already observed in other iron-based compounds~\cite{FeSe, GolubovBaK, HeyerRes}. On the basis of experimental evidences, we assume that the temperature dependence of the resistivity is dominated by the coupling between electrons and spin fluctuations (SF). This simple model allows us to fit very well the resistivity of single crystals by adjusting the free parameters (in particular the characteristic boson frequency $\Omega_0$) in agreement with the results obtained by inelastic neutron scattering and optical measurements~\cite{Inosov,Maksimow}. In the case of thin films, the $\rho(T)$ curves can be fitted with a set of parameters that agree with the experimental data available in literature only if the electron-SF spectrum is depleted at low energy, resulting in a transfer of spectral weight to higher energies. Thanks to the interaction between phonons and spin fluctuations, this effect can be ascribed to the phonon hardening induced by the reduction of the unit-cell volume~\cite{art press} caused by the presence of the CaF$_2$ substrate~\cite{Film1}.

\section{EXPERIMENTAL DETAILS}
The Ba(Fe$_{1-x}$Co$_x$)$_2$As$_2$ ($x = 0.08, 0.10, 0.15$) epitaxial thin films with a thickness of the order of 50 nm were deposited on (001)CaF$_2$ substrates by pulsed laser deposition (PLD)~\cite{Film1} using a polycrystalline target with high phase purity~\cite{Film1,Film2}.
The surface smoothness was confirmed by in-situ reflection high energy electron diffraction (RHEED) during the deposition; only streaky pattern were observed for all films, indicative of a smooth surface. The details of the structural characterization and of the microstructure of these high-quality epitaxial thin films can be found in Ref.~\cite{Film1}. Standard four-probe resistance measurements in van der Pauw configuration were performed in a $^4$He cryostat to determine the $\rho(T)$ curve as well as the transport critical temperature and the width of the superconducting transition \cite{Paola} (both reported in TABLE~\ref{tab:exp}).

\begin{table}
\begin{center}
\begin{tabular}{|c|c|c|c|}
	\hline
 	\%Co &$T\ped{c}\apex{10}$ (K)&$T\ped{c}\apex{90}$ (K)&$\Delta T\ped{c}$ (K)\\
		\hline
	0.08 (TF)& 24.2 &25.6 & 1.4 \\
	0.10 (TF)& 24.6 &26.6 & 2.0\\
	0.15 (TF)& 23.3 &24.8 & 1.5 \\
	0.10 (SC)& 21.4 &22.8 & 1.4\\
		\hline
\end{tabular}
\caption{Critical temperatures of thin films (TF)~\cite{Paola} and single crystals (SC)~\cite{MisureSC} determined from electric transport measurements. $T\ped{c}\apex{10}$ and $T\ped{c}\apex{90}$ are the temperatures
at which the resistance (the resistivity) is respectively 10\% and 90\% of the normal-state value immediately before the transition. $\Delta T\ped{c}$ is defined here as $T\ped{c}\apex{90}-T\ped{c}\apex{10}$.}\label{tab:exp}
\end{center}
\end{table}

\section{THE MODEL FOR THE RESISTIVITY IN A MULTIBAND METAL}
A saturation at high temperature in the normal-state electrical resistivity has been observed in many alloys~\cite{Fisk,Gurvitch} since the 60s. This behavior can be explained within a phenomenological model containing two kinds of carriers with different scattering parameters~\cite{Wiesmann}, then two parallel conductivity channels have to be considered so that
\[
\frac{1}{\rho(T)}=\frac{1}{\rho\ped{{e}}(T)}+\frac{1}{\rho\ped{sat}},
\label{eq:resist2channels}
\]
where $\rho\ped{e}(T)$ is the resistivity of the first group of carriers, characterized by a strong temperature-dependent scattering because of its weak scattering on defects, and $\rho\ped{sat}$ is the contribution of the second group of carriers that gives a strong temperature-independent contribution.

It has been discussed in Ref.~\cite{GolubovBaK} that this shunt model can be derived for hole doped iron pnictides and can explain the normal-state resistivity saturation in Ba$_{1-x}$K$_x$Fe$_2$As$_2$ single crystals.

The resistivity in a multiband case can thus be obtained, extending the single-band case~\cite{Allen, Grimvall} and considering the contribution of all the different channels:
\begin{equation}
\frac{1}{\rho(T)}=\frac{\varepsilon_0}{\hbar}\sum_{i=1}^N \frac{(\hbar\omega_{p,i})^2}{\gamma_i+W_i(T)}, \label{eq:resist}
\end{equation}
where $N$ is the total number of the different carriers considered, $\omega_{p,i}$ is the bare plasma frequency of the $i$th-band and
\begin{equation}
W_i(T)=4\pi k_BT\int_0^\infty d\Omega 			 \left[\frac{\hbar\Omega/2k_BT}{\sinh\big(\hbar\Omega/2k_BT\big)}\right]^2 \frac{\alpha_{tr,i}^2F_{tr}(\Omega)}{\Omega},
\label{W1}
\end{equation}
with $\gamma_{i}=\sum_{j=1}^N(\Gamma_{ij}+\Gamma^{M}_{ij})$, that
is the sum of the inter- and intra-band non-magnetic and magnetic
impurity scattering rates, and
\begin{equation}
\alpha_{tr,i}^{2}F_{tr}(\Omega)=\sum_{j=1}^N\alpha_{tr,ij}^{2}F_{tr}(\Omega),
\end{equation}
where $\alpha^2_{tr,ij}(\Omega)F_{tr}(\Omega)$ are the inter- and intraband  transport electron-boson spectral functions related to the Eliashberg functions~\cite{Allen}.
Just for practical purposes we can define a normalized spectral function $\alpha_{tr,ij}^{2}F'_{tr}(\Omega)$ such that
$\alpha_{tr,ij}^{2}F_{tr}(\Omega)=\lambda_{tr,ij}\alpha_{tr,ij}^{2}F'_{tr}(\Omega)$, where the coupling constants $\lambda_{tr,ij}$ are defined as in Eliashberg theory~\cite{Allen}.

In order to capture the main concepts of the physical problem and not to get lost in a huge number of free parameters, we set all the normalized spectral functions to be equal, i.e. $\alpha_{tr,ij}^{2}F'_{tr}(\Omega)=\alpha_{tr}^{2}F'_{tr}(\Omega)$. In this way the transport spectral functions  $\alpha_{tr,ij}^{2}F_{tr}(\Omega)$ differ only for a scaling factor, i.e. the coupling constant. Incidentally, this is a very good approximation, especially if the coupling is mediated mainly by spin fluctuations, hence suitable for ``122" system. Consequently,
\begin{equation}
\alpha_{tr,i}^{2}F_{tr}(\Omega)=\lambda_{tr,i}\alpha_{tr}^{2}F'_{tr}(\Omega),
\end{equation}
where, obviously, $\lambda_{tr,i}= \sum_{j=1}^N \lambda_{tr,ij}$.  It is also possible to define the total transport coupling constant $\lambda_{tr,tot}=\sum_{i=1}^N N_i\lambda_{tr,i}/\sum_{i=1}^N N_i$ ($N_i$ being the density of the states at the Fermi level of the $i-$th band) for similarity with the superconducting state where $\lambda_{sup,tot}=\sum_{i,j=1}^N N_i\lambda_{sup,ij}/\sum_{i=1}^N N_i$. Note that the specific shape of the spectral function depends on which is the boson that mediates the interaction.

\section{REDUCTION OF A MULTIBAND MODEL TO A TWO-BAND MODEL}
\begin{figure}
    \begin{center}
    \includegraphics[width=0.8\textwidth]{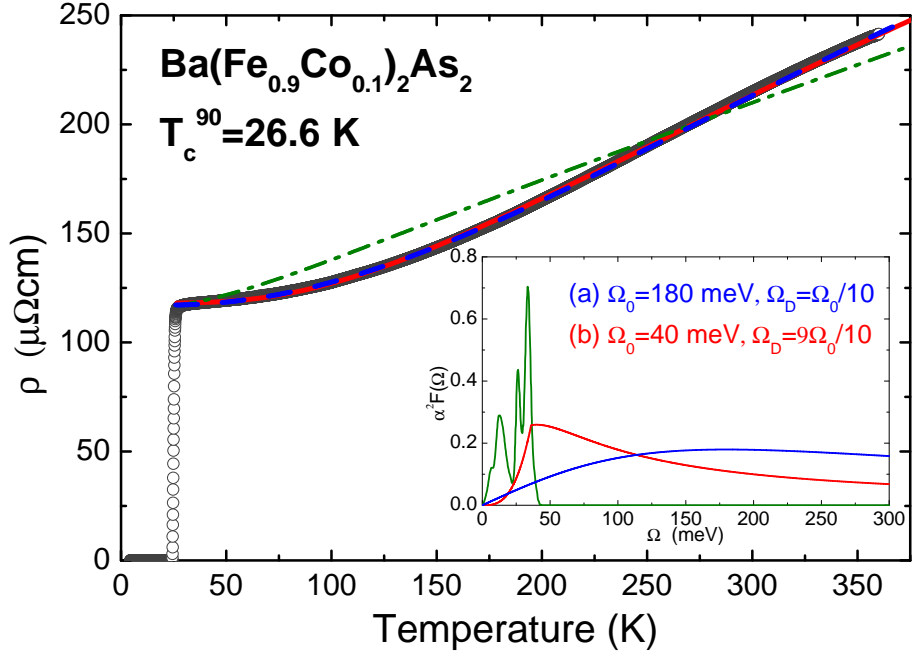}
    \end{center}
    \caption{Temperature dependence of the resistivity of the thin film with $x$=0.10. The dash-dotted green line is the curve obtained within a model containing only one kind of carriers and a phonon mediated coupling. The dashed blue curve represents the fit obtained with $\Omega_{0}=180$ meV and $\Omega_D=0.1 \Omega_0$ (a); the solid red line is the fit obtained with $\Omega_{0}=40$ meV and $\Omega_D= 0.9 \Omega_0$ (b). The relevant normalized electron-boson spectral functions are shown in the inset with the corresponding color.}
    \label{Fig:Co10film}
\end{figure}

Since the Fermi surface of Ba(Fe$_{1-x}$Co$_x$)$_2$As$_2$ presents several sheets, at least in principle a multi-band model should be necessary to explain superconducting and normal-state properties. However one can wonder what is the minimum number of bands to be considered. In the superconducting state, most of the experiments show only two gaps, but in Eliashberg theory a two-band model is not enough to explain the experimental data and the hypothesis of more than two gaps is necessary (the observation of only two distinct gaps can thus be explained by experimental resolution limits and by the similar amplitudes of some of these gaps). Three- and four-band Eliashberg models have been proposed to describe the superconducting phenomenology. However a four-band model contains a huge number of free parameters, and in the present case (in contrast with the case of LiFeAs~\cite{LiFeAs}) it is too hard to fix them in a unique way even if the four-band model is reduced to a simpler effective two-band one~\cite{CharnukhaBaK}.\\
In the normal state, in contrast, the situation appears to be simpler. In order to investigate the electrical resistivity, we group the hole and the electron bands, and we propose a model containing only two different kinds of carriers. Incidentally, that this is the \emph{minimum} model is witnessed by the fact that a single-band model is unable to fit the experimental $\rho(T)$ curve, as shown in FIG.~\ref{Fig:Co10film} \footnote{Owing to the symmetry of the superconducting state, in the single-band model the coupling should be mediated by phonons, whose spectral function is reported in the inset of FIG.~\ref{Fig:Co10film}}.

Considering the fact that the electron-phonon coupling in all the compounds belonging to the class of iron-based superconductors is weak~\cite{Lilia}, it is logical to consider (as done for LiFeAs~\cite{LiFeAs}) that another mechanism contributes to the transport properties; taking into account the superconducting properties of the iron pnictides, the antiferromagnetic (AFM) spin fluctuations are the best candidate to play the role of the principal actor also in the normal state. The electron-SF transport spectral functions $\alpha^2_{tr}(\Omega)F_{tr}(\Omega)$ are similar to the standard Eliashberg function $\alpha^2(\Omega)F(\Omega)$ ~\cite{Allen}, but for $\Omega\rightarrow 0$ they behave like $\Omega^3$ and not like $\Omega$. Therefore, the condition
$\alpha^2_{tr}(\Omega)F_{tr}(\Omega)\propto\Omega^3$
should be imposed in the range $0 <\Omega< \Omega_{D}$, with (according to Ref.~\cite{Allen}) $\Omega_{D}=\Omega_{0}/10$, where $\Omega_{0}$ is the representative bosonic energy.
Then we will take
\begin{equation}
\alpha^2_{tr}(\Omega)F'_{tr}(\Omega)= A\,\Omega^{3}\Theta(\Omega_{D}-\Omega)
+B\,\left[\frac{\Omega_{0}\Omega}{\Omega^{2}+\Omega_{0}^{2}}\Theta(\Omega_{c}-\Omega)\right]\Theta(\Omega-\Omega_{D})\quad
\label{a2F}
\end{equation}
where $\Theta$ is the Heaviside function, and the constants $A$ and $B$ are fixed by requiring the continuity of the function at $\Omega_{D}$ and the normalization.
The factor in square brackets has the functional form of the theoretical AFM SF spectral function in the normal state~\cite{Popovich}, peaked at $\Omega_{0}$, that reproduces the experimental normal-state dynamical spin susceptibility~\cite{Inosov}; here, $\Omega_{c}$ is a cut-off energy (in these calculations \mbox{$\Omega_{c}=1$ eV}).

%
%
\begin{table*}
\begin{center}
\begin{tabular}{|r|c|c|c|c|}
\multicolumn{5}{c}{\textbf{Ba(Fe$_{0.9}$Co$_{0.1}$)$_{2}$As$_{2}$}}\\
\hline
\multirow{2}{*}{}&\multicolumn{3}{c|}{\textbf{Thin film}}&\textbf{Single Crystal}\\
\cline{2-5}
&1-band model&2-band model (a)& 2-band model (b)& 2-band model\\
\hline
\hline
$\Omega_0$ (meV)& 	-	& 180	& 40 & 40\\
$\lambda_{tr,1}$&		0.20 & 0.65		& 0.33& 0.35\\
$\lambda_{tr,tot}$&		0.20	& 0.26	& 0.13& 0.14\\
$\gamma_1$ (meV)&	37	& 21		& 24&16\\
$\gamma_2$ (meV)&	-	& 69		& 100&80\\
$\omega_{p,1}$ (meV)&	1550& 1060	& 1180&1110\\
$\omega_{p,2}$ (meV)&	-	& 1021	& 1020&700\\
\hline
\end{tabular}
\caption{ Values of the parameters used for the fit of the resistivity of the Ba(Fe$_{0.9}$Co$_{0.1}$)$_{2}$As$_{2}$ thin film reported in FIG.~\ref{Fig:Co10film} and for the fit of the resistivity of the single crystal reported in FIG.~\ref{Fig:Co10comp}. In all the cases $\lambda_{tr,2}=0$, so that $\lambda_{tr,tot} = (N_1\lambda_{tr,1})/(N_1+N_2)$. $\gamma_1$, $\gamma_2$, $\omega_{p,1}$, $\omega_{p,2}$ are not independent parameters because they are related to $\rho_0$ by eq. \ref{eq:resist}.}\label{tab:Co10}
\end{center}
\end{table*}
\begin{figure}
    \begin{center}
            \includegraphics[width=0.8\textwidth]{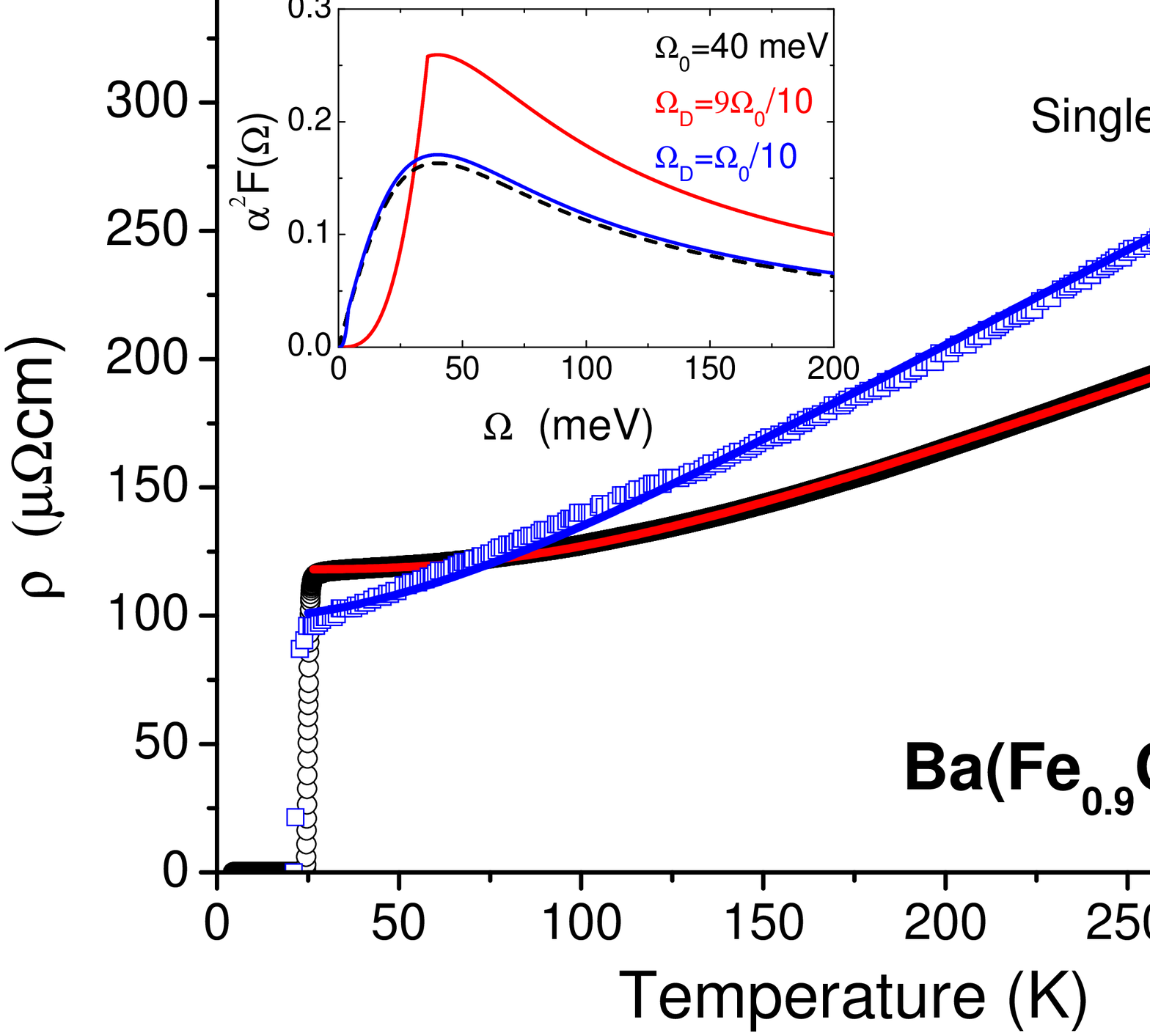}
    \end{center}
    \caption{Comparison between the resistivity of Ba(Fe$_{0.9}$Co$_{0.1}$)$_2$As$_2$ thin film (black circles and solid red line) and single crystal (blue squares, taken from \cite{MisureSC}, and solid blue line). The fit (line) has been obtained in both cases with \mbox{$\Omega_{0}=40$~meV} taking $\Omega_D=0.1 \Omega_0$ for the single crystal, and $\Omega_D=0.9 \Omega_0$ for the thin film. The inset shows the two spectral functions that have been used to fit the experimental data with the corresponding colors and the antiferromagnetic spin fluctuations spectral function in the normal state (dashed black line) without the change of behavior at low energy.}
    \label{Fig:Co10comp}
\end{figure}

As mentioned above, in these compounds the electronic structure consists of electron and  hole pockets (hereafter indicated respectively by the indices 1 and 2). Then, in order to keep the number of free parameters as low as possible and taking into account that just one kind of carriers is not enough,  we consider a model containing two different kinds of carriers. Within this model the electron-boson coupling constants $\lambda_{tr,1}$ and $\lambda_{tr,2}$, the impurities scattering rates $\gamma_1$ and $\gamma_2$, the plasma energies $\omega_{p,1}$ and $\omega_{p,2}$ and the representative energy $\Omega_{0}$ of the transport electron-boson spectral functions are the free parameters. Specific properties of each compound and experimental data allow to fix some of them. For instance, $\Omega_{0}$ can be derived from inelastic neutron scattering experiments, while the values of $\omega_{p,1}$ and $\omega_{p,2}$ can be obtained, at least in the case of ideal single crystals, from first principle calculations. Moreover, ARPES and de Haas-van Alphen data suggest that for Co-doped Ba-122 the transport is dominated by the electronic bands and that the hole bands are characterized  by a smaller mobility~\cite{MisureSC, Maksimow}. This means that within our model: (i) the transport coupling is much stronger in the electron band, so that, at least as a first approximation, $\lambda_{tr,2}$ can be fixed to zero and (ii) the effects of the impurities are mostly concentrated in the hole band, i.e. $\gamma_{2}>\gamma_{1}$.
In this way one contribution to the resistivity (the hole one) results to be temperature independent, as can be noticed by looking at equations \eref{eq:resist} and \eref{W1}, and the other is responsible for the slope of the resistivity with the temperature. Moreover the values of $\gamma_2$ and $\gamma_1$ are related to the value of the residual resistivity $\rho_0$ through the plasma frequencies, therefore the number of the degrees of freedom is decreased by one.
In the best case, this simplified model contains only two free parameters ($\lambda_{tr,1}$ and, for example, $\gamma_2$). However, in the case of Ba(Fe$_{1-x}$Co$_{x}$)$_{2}$As$_{2}$ thin films there are not enough experimental data to fix $\Omega_{0}$ in the normal state, which is expected to change with the cobalt content. On the other hand, the combination of various factors such as the doping homogeneity, the strain due to the substrate and so on, make ab-initio calculations of $\omega_{p,1}$ and $\omega_{p,2}$ somehow unreliable and arbitrary. Therefore, we keep $\omega_{p,1}$ and $\omega_{p,2}$ as adjustable parameters, but we assume them not to deviate very much from the values determined in the optimal-doping case in the same films~\cite{Maksimow} (\mbox{$\omega_{p,1}\thicksim 1.1$ eV} and \mbox{$\omega_{p,2}\thicksim$ 0.7 eV}). Similarly, $\Omega_{0}$ is allowed to change, but compatibly with the value determined from inelastic neutron scattering measurements in optimally-doped Ba(Fe$_{1-x}$Co$_{x}$)$_{2}$As$_{2}$ (i.e. $\Omega_{0}$=40~meV \cite{Inosov}).
These constraints, combined with the condition on $\gamma_i$ and the particular shape of the experimental resistivity as a function of temperature, strongly limit the range of parameter values that allow fitting a given $\rho(T)$ curve, so that only small variations of each parameter around its best-fitting value are allowed.
\section{RESULTS AND DISCUSSION}
The first step was to apply this model to the resistivity of the Ba(Fe$_{1-x}$Co$_x$)$_2$As$_2$ thin film with 10\% of cobalt (see FIG.~\ref{Fig:Co10film}). If the spectral functions are kept with the form of equation~\eref{a2F}, the peak energy $\Omega_0$ is fixed as explained above, and $\Omega_D = \Omega_0 /10$, there is no way to adjust the other input parameters to reproduce the experimental $\rho(T)$ curve. This is true even if the plasma frequencies take very different values with respect to those mentioned above.

In contrast, with this input we can reproduce very well the resistivity of the single crystals with the same doping, as shown in FIG.~\ref{Fig:Co10comp}.
To reproduce the low-temperature behavior of the $\rho(T)$ curve in thin films, we are forced to act on the shape of $\alpha^2_{tr}F'_{tr}(\Omega)$.  If its functional form is kept as in equation \eref{a2F} with $\Omega_D = \Omega_0 /10$, the representative bosonic energy has to be drastically increased up to $\Omega_{0}$=180~meV (as shown in FIG.~\ref{Fig:Co10film}). Such a high value of the energy of the mediating boson looks unreasonable~\cite{Inosov}. The only other possibility is to keep $\Omega_0$ to the same value used in single crystals, and to change the range where the spectral function follows a $\Omega^3$ trend, i.e. to increase $\Omega_D$ well above the usual maximum value of $\Omega_0$/10.  This assumption allows a very good fit of the resistivity of 10\% Co-doped thin films, as shown in FIG.~\ref{Fig:Co10film} and in FIG.~\ref{Fig:Co10comp} (solid red lines). All the parameters that have been used to fit the resistivity of 10\% Co-doped Ba-122 are reported in TABLE~\ref{tab:Co10} (column b for the thin film and last column for the single crystal).

\begin{figure}
    \begin{center}
    \includegraphics[width=0.8\textwidth]{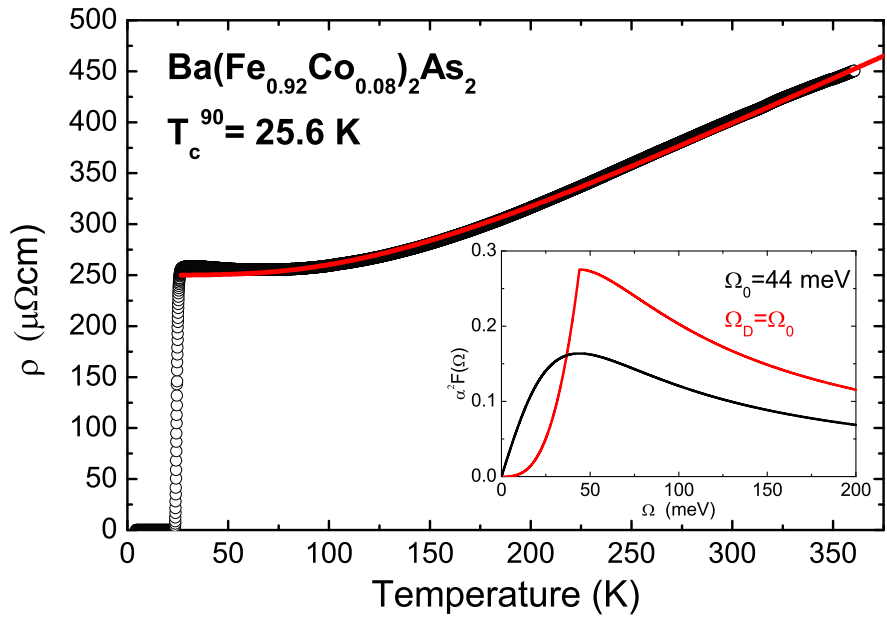}
    \end{center}
    \caption{Temperature dependence of the resistivity of the Ba(Fe$_{1-x}$Co$_x$)$_2$As$_2$ thin film with $x=0.08$. The solid red curve represents the fit obtained with $\Omega_{0}=\Omega_D=44$ meV. In the inset two different electron-boson spectral functions are reported: the red curve is that used to fit the  experimental data, the black one is the antiferromagnetic spin fluctuations spectral function in the normal state without changing the low-energy behavior.}
    \label{Fig:Co8film}
\end{figure}
\begin{figure}
    \begin{center}
    \includegraphics[width=0.8\textwidth]{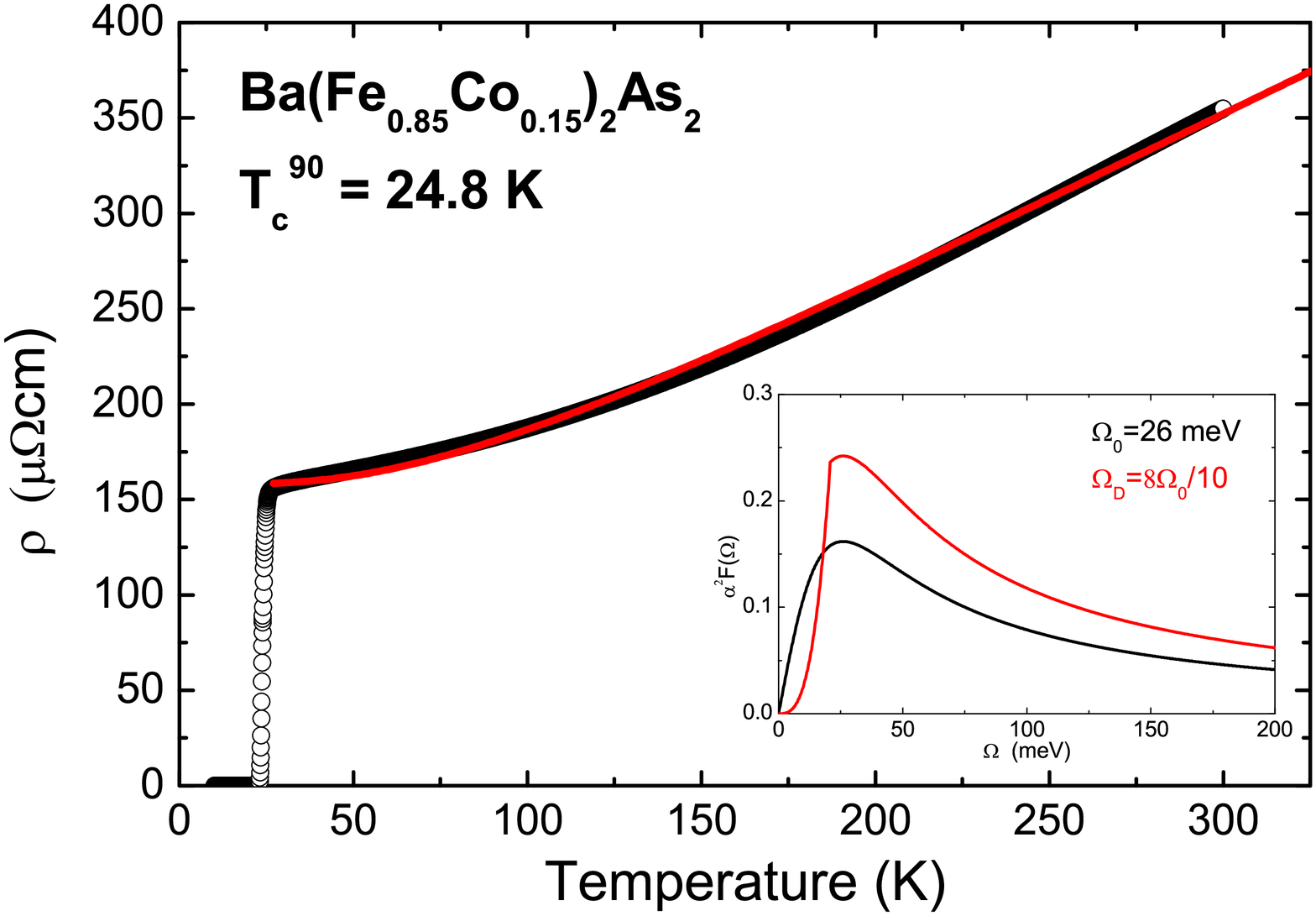}
    \end{center}
    \caption{Temperature dependence of the resistivity of the Ba(Fe$_{1-x}$Co$_x$)$_2$As$_2$ thin film with $x$=0.15. The solid red curve represents the fit obtained with $\Omega_{0}=26$ meV and $\Omega_D= 0.8 \Omega_0$. In the inset two different electron-boson spectral functions are shown: the red curve is the one used to fit the experimental data, the black one is the antiferromagnetic spin fluctuations spectral function in the normal state without changing the low-energy behavior.}
    \label{Fig:Co15film}
\end{figure}

An analogous procedure has been applied also to Ba(Fe$_{0.92}$Co$_{0.08}$)$_2$As$_2$ and to Ba(Fe$_{0.75}$Co$_{0.15}$)$_2$As$_2$ thin films as shown in FIG.~\ref{Fig:Co8film} and FIG.~\ref{Fig:Co15film}. Also in these cases the resistivity was initially reproduced with an unphysically high bosonic peak energy and then, following the same reasoning, the shape of the spectral function was changed and a good fit was obtained with the values reported in TABLE~\ref{tab:Co8}. This leads to similar conclusions and supports the idea that in the case of thin films the spectral function should have a $\Omega^3$ behavior in a wider range than in single crystals. In other words, there is a ``hardening'' of the electron-boson spectral function, with a transfer of spectral weight from energies smaller than $\Omega_0$ to energies higher than $\Omega_0$. Probably this effect is so striking because the main mechanism is not phononic. In fact the strain produced by the substrate on the film can create changes in the electronic structure and in the Fermi surface nesting that is crucial for raising superconductivity mediated by spin fluctuations~\cite{Anna}. The effect of the substrate can be assimilated to that of a uniaxial pressure and in particular~\cite{KazuTcStrain} the substrate used here (CaF$\ped{2}$) has the largest effect on the volume cell of the Co doped Ba-122 because of the mismatch of the dimension of the unit cells.
From a qualitative point of view we know that even if for the ``122'' family the phonon coupling is small and basically negligible at zero pressure, a reduction of the unit cell volume causes an increase of phonon frequencies especially for the superconducting compounds~\cite{art press}.
Concerning the parent compound,  it has been argued that under pressure the interaction between the lattice structure and the magnetism is no longer negligible even in the normal state~\cite{NatmatPhSp}, i.e the spin-phonon coupling may play an important role. The central point is that the $A\ped{1g}$ vibration of the arsenic ions slightly distort the AsFe$\ped4$ tetrahedron and this has effects even in the normal state. This could cause a ``hardening'' of the transport electron-SF spectral function.

Incidentally, it is interesting to note, as shown in TABLE~\ref{tab:Comparison}, that the representative energy of the electron-boson spectral function in the normal state decreases with increasing the cobalt content in the Ba(Fe$_{1-x}$Co$_x$)$_2$As$_2$ thin films. The same table also highlights that in this compound (no matter if in the form of crystal or thin film) as well as in other iron-based superconductors~\cite{GolubovBaK,LiFeAs} the normal and superconducting states are characterized by very different values of the typical energy of the electron-boson spectral function and of the total electron-boson coupling constant. In particular, the typical energy of the electron-boson spectral function systematically increases going from the superconducting to the normal state (in agreement with inelastic neutron scattering experimental data~\cite{Inosov}) while the total electron-boson coupling constant significantly decreases. The fact that the same property has been observed in HTCS~\cite{Maxi} may suggest that iron-based compounds share some characteristics with cuprates.

\begin{table}
\begin{center}
\begin{tabular}{|r|c|c|c|}
\multicolumn{4}{c}{\textbf{Ba(Fe$_{1-x}$Co$_{x}$)$_{2}$As$_{2}$ thin films }}\\
\hline
			&$x$=0.08		& $x$=0.1		&$ x$=0.15\\
		\hline
		\hline
		$\Omega_0$ (meV)	& 	 44		& 40 & 26\\
        $\Omega_D$ (meV)		& 	 $\Omega_0$	& $\;9\Omega_0/10\;$ & $\;8\Omega_0/10\;$\\
		$\lambda_{tr,1}$&		 0.32	& 0.33& 0.34\\
		$\lambda_{tr,tot}$&		 0.13& 0.13& 0.14\\
		$\gamma_1$ (meV)&	 32		& 24& 26\\
		$\gamma_2$ (meV)&	 190		& 100& 600\\
		$\omega_{p,1}$ (meV)&	 970	& 1180& 1140\\
		$\omega_{p,2}$ (meV)&	 780	& 1020& 1140\\
		\hline
\end{tabular}
\caption{Values of the parameters used for the fit of the $\rho(T)$ curves of Ba(Fe$_{1-x}$Co$_{x}$)$_{2}$As$_{2}$ thin films with three different Co contents ($x$ = 0.08, 0.10, 0.15). In all the cases $\lambda_{tr,2}=0$.}\label{tab:Co8}
\end{center}
\end{table}
\begin{table}
\begin{center}
\begin{tabular}{|r|c|c|c|c|c|}
\hline
	& T$\ped{c}$	&$\Omega_{0}$	&$\Omega_{0}^{sup}$	&$\lambda_{tr,tot}$	&$\lambda_{sup,tot}$\\
\hline	
\hline				   	
\footnotesize{LiFeAs}						            &18	    & 47	&  8	& 0.77   &2.00\\
\footnotesize{Ba(Fe$_{0.9}$Co$_{0.1}$)$_2$As$_2$}       &22.8	& 40	&  9	& 0.14   &2.83\\
\footnotesize{TF-Ba(Fe$_{0.92}$Co$_{0.08}$)$_2$As$_2$ } &25.5	& 44	& 10	& 0.13   &2.22\\
\footnotesize{TF-Ba(Fe$_{0.9}$Co$_{0.1}$)$_2$As$_2$ }   &26.6	& 40	& 11   	& 0.13   &2.22\\
\footnotesize{TF-Ba(Fe$_{0.75}$Co$_{0.15}$)$_2$As$_2$ } &24.8	& 26	&  9	& 0.14   &1.82\\
\footnotesize{Ba$_{0.68}$K$_{0.32}$Fe$_2$As$_2$ }       &38.5	& 40	& 18	& 0.87   &1.89\\
\hline
\end{tabular}
\caption{Summary of characteristic parameters obtained in this work for Ba(Fe$_{1-x}$Co$_{x}$)$_2$As$_2$ thin films (TF) and single crystals~\cite{MisureSC}, in a similar work done on LiFeAs~\cite{LiFeAs} and taken from literature for the case of Ba$_{0.68}$K$_{0.32}$Fe$_2$As$_2$~\cite{GolubovBaK}. The values of $T\ped{c}$, of the characteristic spin-fluctuation energy (in the normal and in the superconducting state, $\Omega_0$  and $\Omega_{0}^{sup}$), and of the total coupling constant (in the normal and in the superconducting state, $\lambda_{tr,tot}$ and $\lambda_{sup, tot}$) are shown.
The critical temperatures are expressed in kelvin and the energies in meV.}\label{tab:Comparison}
\end{center}
\end{table}
\section{CONCLUSIONS}
It is already known that, in thin films of iron-based compounds, the transport critical temperature is affected by the strain due to the substrate \cite{KazuTcStrain}. Here we have shown that this effect may also explain the different shape of the $\rho(T)$ curve in crystals and thin films of Co-doped Ba-122, even when the Co content is the same. By using a simple effective two-band model for the transport, we have shown that the $\rho(T)$ curves of thin films can be fitted with the same characteristic boson energy $\Omega_0$ used for single crystals only if some spectral weight is transferred from below $\Omega_0$ to above $\Omega_0$ in the transport electron-boson spectral function $\alpha^2_{tr}F_{tr}(\Omega)$. In other words, $\alpha^2_{tr}F_{tr}(\Omega)$ behaves like $\Omega^3$ in a wider energy range than in single crystals. This effect is possibly induced by the indirect coupling between spin fluctuations and phonons \cite{NatmatPhSp}, whose frequencies are increased by the compressive strain \cite{art press}. Finally, we have shown that the decrease (increase) in the electron-boson coupling constant (characteristic boson energy) between the superconducting and the normal state ($\lambda_{sup,tot} \gg \lambda_{tr,tot}$, $\Omega_0^{sup} \ll \Omega_0$) observed in crystals of various Fe-based compounds also holds  in thin films of Co-doped Ba-122. This behavior is also common to HTSC \cite{Maxi} and may thus be a unifying principle at the root of superconductivity in iron-based materials and cuprates.
\section{ACKNOWLEDGMENTS}
The authors would like to thank F. Galanti for his cooperation in resistivity measurements.
This work was done under the Collaborative EU-Japan Project ``IRON-SEA'' (NMP3-SL-2011-283141).\\
%

\end{document}